\documentclass[preprint,aps]{revtex4}
\usepackage{graphicx}

\begin{document}

\title{Spin-Polarized Current Induced Torque in Magnetic Tunnel Junctions}

\author{A.~Kalitsov$^1$, I.~Theodonis$^{1,2}$ and N.~Kioussis$^1$}

\affiliation{$^1$Department of Physics, California State University, Northridge, CA, USA \\ 
$^2$Department of Physics, Ethniko Metsovio Polytechnio, GR-15773, Zografou, Athens, Greece}

\author{M.~Chshiev, W.~H.~Butler}

\affiliation{MINT Center, University of Alabama, P.~O.~Box 870209, Tuscaloosa, AL, USA}

\author{A.~Vedyayev}

\affiliation{Faculty of Physics, M. V. Lomonosov Moscow State University, Moscow, Russian Federation}

\begin{abstract}

We present tight-binding calculations of the spin torque in non-collinear magnetic tunnel junctions based on the non-equilibrium Green functions approach. We have calculated the spin torque via the effective local magnetic moment approach and the divergence of the spin current. We show that both methods are equivalent, i.e. the absorption of the spin current at the interface is equivalent to the exchange interaction between the electron spins and the local magnetization. The transverse components of the spin torque parallel and perpendicular to the interface oscillate with different phase and decay in the ferromagnetic layer (FM) as a 
function of the distance from the interface. The period of oscillations is inversely proportional to the difference between the Fermi-momentum of the majority and minority electrons. The phase difference between the two transverse components of the spin torque is due to the precession of the electron spins around the exchange field in the FM layer. In absence of applied bias and for a relatively thin barrier the perpendicular component of the spin torque to the interface is non-zero due to the exchange 
coupling between the FM layers across the barrier.

\end{abstract}


\date{\today}
\maketitle

The effect of current-induced switching on the orientation of magnetic 
moments in magnetic heterostructures has recently attracted significant 
interest due to its potential applications to spin 
electronics~\cite{Myers,Krivorotov}, such as current-controlled magnetic
memory elements. When a current of spin polarized electrons enters a 
ferromagnet, there is a transfer of angular momentum between the 
propagating electrons and the background magnetization of the layer. 
The concept of switching the orientation of a magnetic layer of a 
multilayered structure by the current perpendicular to the layers was 
originally proposed by Slonczewski~\cite{Slonczewski} and 
Berger~\cite{Berger}, and has been followed by others~\cite{Ralph,Stiles}. 
This effect, often referred to as ``spin torque", may, at sufficiently high 
current densities, alter the magnetization state. An alternative mechanism 
of current-induced switching was put forth by Heide 
{\it et al}~\cite{Heide} and Zhang {\it et al}~\cite{Levy} in which the 
current across the magnetically inhomogeneous multilayer produces spin 
accumulation which establishes an energy preference for a parallel or 
antiparallel alignment of the moments of the magnetic layers. 

The purpose of this work is to employ a one-dimensional single-band 
tight-binding model and calculate the spin torque in magnetic tunnel 
junctions consisting of two ferromagnetic layers (FM) with non-collinear 
orientation of the magnetization, separated by an insulating (I) thin 
layer. The calculations are carried out using the non-equilibrium 
Green's function technique in the framework of the Keldysh 
formalism~\cite{Caroli}. We have calculated the spin torque using two 
different approaches: the first uses the local magnetic moment of the 
conduction electrons and the exchange splitting of the $d$-band; the 
second uses the divergence of the transverse component of the spin current. 
The results demonstrate that these two approaches are equivalent, as 
hypothesized originally by Edwards {\it et al}~\cite{Mathon}. 

The method for calculating the non-equilibrium Green functions for the 
non-collinear system is a generalization of that introduced by Caroli 
{\it et al}~\cite{Caroli} for nonmagnetic leads, where one replaces the 
Green functions with 2$\times$2 matrices in spin space. For this purpose 
we divide the Hamiltonian into two parts: one which describes the 
spin-average medium and the other proportional to the spin splitting 
of the band structure, i.e.,
\begin{equation}
\stackrel-{H}_{pq}=\frac{1}{2} \left ( \epsilon_p^{\uparrow} + 
\epsilon_p^{\downarrow} \right ) + \frac{1}{2} \left ( t_{pq}^{\uparrow} 
+ t_{pq}^{\downarrow} \right ), 
\hspace{ 0.3 in}
\delta H_{pq}=\frac{1}{2} \left ( \epsilon_p^{\uparrow} - 
\epsilon_p^{\downarrow} \right ) + \frac{1}{2} \left ( t_{pq}^{\uparrow} - 
t_{pq}^{\downarrow} \right ).
\end{equation}
Here, the subscript indexes indicate the atomic sites, the superscript 
indexes refer to spins, and $\epsilon^{\uparrow(\downarrow)}$ and 
$t^{\uparrow(\downarrow)}$ denote the spin-dependent on-site energy and 
nearest-neighbor hopping matrix elements, respectively. The one-electron 
Shr\"odinger equation becomes
\begin{equation}
\sum_{p_1} \left [ (E \delta_{pp_1}-\stackrel-{H}_{pp_{1}})I-\delta 
H_{pp_{1}} \left ( \begin{array} {ccc} cos \gamma & sin \gamma \\ 
sin \gamma & -cos \gamma \\ \end{array} \right ) \right ] \left ( 
\begin{array} {ccc} g_{p_{1}q}^{\uparrow \uparrow} & 
g_{p_{1}q}^{\uparrow \downarrow} \\ g_{p_{1}q}^{\downarrow \uparrow} & 
g_{p_{1}q}^{\downarrow \downarrow} \\ \end{array} \right ) = \delta_{pq} I,
\end{equation}
where $g_{pq}^{\sigma_1\sigma_2}$ is the retarded Greens function for the 
isolated semi-infinite leads, $\gamma$ is the angle between the 
magnetizations of the FM leads in the plane perpendicular to the current, 
and $I$ is 2$\times$2 unit matrix. Following Caroli~\cite{Caroli}, we next 
calculated the retarded Green function for the entire system. Finally, 
by solving the kinetic equation we evaluated the non-equilibrium Green function $G^<$.

Having calculated the $G^<$, the effective local magnetic moment on site 
$i$ in the right FM lead can be obtained from it by
\begin{equation}
\mbox{\boldmath$\mu$}_{i} = -\frac {i \mu_B}{2\pi} \int 
Tr_\sigma \left [ G_{ii}^{< \sigma_1, 
\sigma_2} \mbox{\boldmath$\sigma$} \right ] dE,
\end{equation}
where \mbox{\boldmath$\sigma$} is the Pauli matrix vector. The spin current is given by
\begin{equation}
\mbox{\boldmath$Q$}_{i,i+1} = \frac{1}{2} 
\int Tr_\sigma \left [ (G^{< \sigma_{1},\sigma_{2}}_{i,i+1} - 
G^{< \sigma_1, \sigma_2}_{i+1,i}) 
T^{\sigma_{1}, \sigma_{2}} \mbox{\boldmath$\sigma$} \right ] dE,
\end{equation}
where $T^{\sigma_{1}, \sigma_{2}}$ is the hopping matrix element in the right lead. 

We model the FM/I/FM junction with $\epsilon^{\uparrow}=0.794 t^\uparrow$, 
$\epsilon^{\downarrow}=1.794 t^\downarrow$, 
$t^{\uparrow}=t^{\downarrow}=0.4~e\mathrm{V}$ for the FM leads, and 
$\epsilon=3 t_b$, $t_b=1~e\mathrm{V}$ for the insulating 
layer~\cite{Mathon2}. The insulator consists of five layers. We assume that the potential drop
is confined to the barrier and is linear within the barrier.  We have 
calculated the torque from
\begin{equation}
\mbox{\boldmath$T$}=\mbox{\boldmath$\Delta$} \times \mbox{\boldmath$\mu$},
\end{equation}
where ${\bf \Delta} = \hat z(\epsilon^{\uparrow}-\epsilon^{\downarrow})/2\mu_B$ 
is the exchange field along the z-direction, and, alternatively, from the divergence of 
the spin current
\begin{equation}
{\bf T}=-\nabla\cdot{\bf Q}.
\end{equation}

In Fig.1 we show the spatial distribution of the components of the spin torque on 
the right FM lead that are perpendicular ($T_\bot$) and parallel ($T_{||}$) to the plane of the layers
for $\gamma=\pi/2$.  These are calculated both via the local magnetic moment approach 
using equation (5) (solid and dashed curves) and via the divergence of the 
spin current from equation (6) (solid and open symbols). 
The direction of eletron flow is from the right to the left FM lead.
\begin{figure}
\begin{center}
\includegraphics[width=7.76cm]{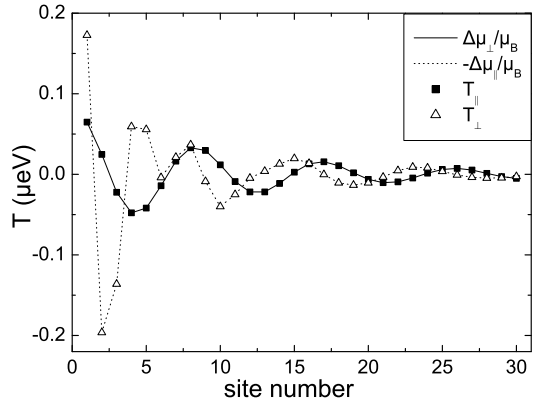}
\caption{Site-dependence of the  perpendicular ($T_{\bot}$) and parallel 
($T_{||}$) components of the spin torque on the right FM lead calculated 
via the local magnetic moment approach (Eq. (6))and the divergence of the 
spin current (Eq. (7)), for $\gamma=\pi / 2$ and $V=0.1$~V}
\end{center}
\end{figure}
One can see that both methods are equivalent giving identical results for 
the spin torque, i.~e. the absorption of the spin current at the right FM electrode  
is equivalent to the exchange interaction between the electron 
spins and the local magnetization.  This equivalence is valid for an arbitrary 
angle $\gamma$ between the magnetization of FM layers. The angular dependence 
of the spin torque is proportional to $\sin \gamma$.

Both $T_{\bot}$ and $T_{||}$ components of the spin torque oscillate with different phase 
and decay with distance in the FM layer from the right FM electrode. The 
period of oscillations is inversely proportional to the difference 
between the Fermi-momentum of spin-up and spin-down electrons. The phase 
difference between the two components of the spin torque is due to the 
precession of the electron spins around the exchange field in the FM layer.
The components $\mu_{||}$ and $\mu_{\bot}$ of the effective magnetic moment 
are referred to by previous studies\cite{Heide,Levy} as ``spin accumulation".
We find that in general, $T_{\bot}$ and $T_{||}$ are 
comparable in magnitude and they can be of the same or opposite sign depending 
on the position in the right FM electrode. 

In Fig.~2 we show the spatial distribution of the parallel and 
perpendicular components of the spin torque at 
zero and finite bias. At zero bias, $T_{||}=0$ while $T_{\bot}\neq0$, 
indicating that $T_{\bot}$ plays the role of the exchange coupling between 
the FM leads across the barrier~\cite{Tiusan}. At finite bias, the role of current-induced 
spin torque is played by $T_{||}$ and $T_{\bot}(V\neq 0)-T_{\bot}(V=0)$.
The exchange coupling explains irregular oscillations of $T_{\bot}$ in Fig.~1 
and Fig.~2 at finite bias close to the interface. Since it decays rapidly 
with the distance from the interface~(see Fig.~2, $T_{\bot}$ at $V=0$), the 
oscillatory behaviour of $T_{\bot}$ becomes normal again in the bulk of 
the FM layer. 

\begin{figure}
\begin{center}
\includegraphics[width=7.76cm]{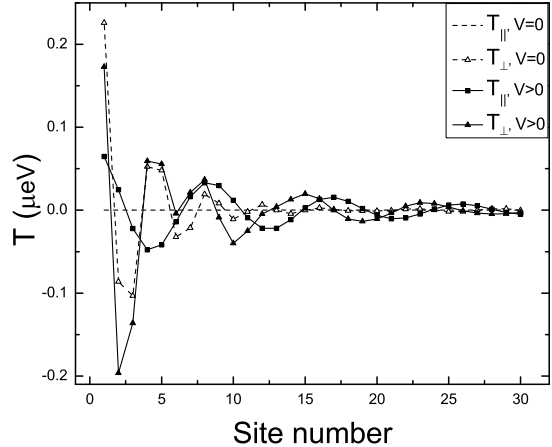}
\caption{Spatial distribution of the parallel and perpendicular components of 
the spin torque at zero and 0.1~V bias for $\gamma=\pi/2$.}
\end{center}
\end{figure}

So far we have discussed the spin torque behaviour for the case of positive applied bias, 
i.e. when electrons flow from the right FM electrode to the left one. We
consider the spin torque in the right FM electrode, where the magnetization orientation 
was chozen to be along the $\hat z$ axis. In this case, only those electrons reflected from the left 
interface are polarized transversly to the magnetization of the right FM layer 
thereby contributing to the current induced spin torque. The situation changes drastically 
when one reverses the current flow, i.e. at negative bias. The majority of
electrons transmitted from the left FM layer into the right one are transversely polarized 
and lead to a significant increase of  the current induced spin torque. 
Both situations are demonstrated in Fig.~3, where $T_{||}$ and $T_{\bot}$ are plotted 
for positive and negative applied bias.

\begin{figure}
\begin{center}
\includegraphics[width=7.76cm]{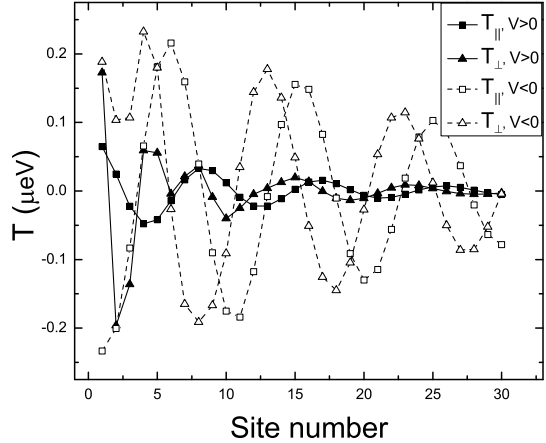}
\caption{Spatial distribution of the parallel and perpendicular components of 
the spin torque for positive ($V=0.1$~V) and negative bias ($V=-0.1$~V).}
\end{center}
\end{figure}

The research at California State University, Northridge was supported from NSF
under Grant No DMR-00116566 and US Army under Grant No AMSRD-45815-PH-H .

\pagebreak

%
%
%

\end{document}